\documentclass[conference,compsoc]{IEEEtran}
\usepackage{graphicx,mathtools,amssymb,algpseudocode,algorithm,url}

\ifCLASSOPTIONcompsoc
  \usepackage[nocompress]{cite}
\else
  \usepackage{cite}
\fi
\ifCLASSINFOpdf
\else
\fi

\usepackage{caption} 
\begin{document}
%
\title{Multi-Agent Actor-Critic in Autonomous Cyber Defense}

\author{\IEEEauthorblockN{Mingjun Wang}
\IEEEauthorblockA{World Wide Technology\\
Email: mingjunwang88@gmail.com}
\and
\IEEEauthorblockN{Remington Dechene}
\IEEEauthorblockA{World Wide Technology\\
Email: Remington.Dechene@wwt.com}}


%


\maketitle

\begin{abstract}
 As cyber threats continue to grow in complexity and frequency, the necessity for autonomous, adaptive, and intelligent defense mechanisms has become increasingly urgent. In this work, we examine the application of Multi-Agent Deep Reinforcement Learning (MADRL) to the domain of autonomous cyber defense, with a particular focus on Actor-Critic algorithms. These algorithms offer a general and scalable framework for multi-agent learning and coordination, making them well-suited for complex, dynamic cybersecurity environments. In this paper, we focus on cooperative MADRL-based systems in which case multiple agents operate collaboratively to detect, mitigate, and respond to cyber threats in real time. By leveraging shared experiences and coordinated learning, each agent independently acquires policies that enable rapid adaptation to the attack patterns. We validate the effectiveness of this approach through simulated cyber-attack scenarios designed to evaluate the system’s responsiveness, adaptability, and robustness. The experimental results demonstrate that agents trained via the MADRL framework can autonomously learn to counteract diverse threats with high efficiency. This not only enhances the operational resilience of cyber defense systems but also reduces reliance on manual oversight. Our findings suggest that MADRL holds considerable promise for advancing the state of the art in autonomous cyber defense, offering a pathway toward more intelligent, scalable, and resilient cybersecurity architectures. This paper provides a comprehensive view of Actor-Critic in coorpertive MADRL and its application in cybersecurity. It contributes to the growing body of research at the intersection of artificial intelligence and cybersecurity and provides a foundation for future exploration into decentralized, learning-based defense strategies.
\\\\Keywords: Deep Learning, Multi-Agent Reinforcement Learning, Cybersecurity. 
\end{abstract}


%
\IEEEpeerreviewmaketitle

\section{Introduction}
Network security is characterized by significant asymmetry, as defenders must ensure the continuous protection of the network's components from various attackers, while adversaries need only exploit a single weak entry point at any given time. This generates a great demand for continuous automated cyber-defense systems. The majority of current security techniques are classified as an intrusion detection system(IDS). The early generations of IDS were predominantly rule-based. However, recent advancements have seen the application of machine learning and deep learning in cyber-defense. These models utilize traffic features such as packet header information to predict the likelihood of a traffic being a threat. Unfortunately, such models often suffer from high false positive rate. 

Reinforcement learning (RL) excels in sequential interactive decision making \cite{sutton2018reinforcement} that cannot be easily solved using analytical solutions. RL algorithms with mindset of optimizing long-term goals. This makes it different from the classical supervised learning based models. With the help of deep learning, Deep Reinforcement leaning(DRL) is able to tackle large observation and action spaces. DRL trained agents achieved human and even superhuman levels of performance in a range of complex tasks including classic board games such as Go \cite{silver2017mastering}, video games ranging from classic Atari \cite{mnih2013playing}, autonomous driving \cite{sallab2017deep}, and robotics \cite{ibarz2021train}. DRL has also been successfully applied to autonomous network defense \cite{foley2023inroads}, a task in which the defend agent proactively monitors the state of the network, identifies abnormalities, and takes action to remediate.

In a more realistic word, security engineers face large enterprise networking systems that often comprise different segments. These segments may have different operational systems and objectives. A single agent would not be able to tackle these complicated networks. Multi-Agent system appears to be a good choice. In this paper, we investigate to utilize MADRL in a cooperative stetting. More specifically we assume each agent shares a common reward. Their common goal is to maximize the long term total rewards. We will explore several variants of Multi-Agent Actor-Critic algorithms in autonomous cyber defense system. 

\section{Background}
\subsection{Actor Critic Algorithms}
Actor-Critic (AC) algorithms are a class of reinforcement learning methods that combine the benefits of both policy-based and value-based approaches. They utilize two primary components: the actor and the critic. Actor maps states to a probability distribution over action space. Critic measures the expected return (future rewards) of being in a given state or state-action pair under policy $\pi$. The actor is updated under policy gradient methods maximizing the expected total return by repeatedly estimating the gradient $g:=\triangledown_{\theta}\sum_{t=0}^{T}\gamma_{t}r_{t}$

The general formalism for Actor-Critic algorithms is based on policy gradient from \cite{sutton2018reinforcement}

\begin{equation}\label{eq:1}
    g_t:=\mathbb{E} [\triangledown_{\theta}\log\pi_\theta(a_t | s_t)A_t]
\end{equation}
\[
where \quad A_{t} = Q_{\pi}(s_{s},a_{t}) - V_{\pi}(s_{t})
\]
$A_t$ is advantage function measured by the difference between state-action value and state value. $Q_{\pi}(s_{t},a_{t})$ can be estimated as $r_{t} + V_{\pi}(s_{t},a_{t})$  More advanced GAE was proposed by \cite{schulman2015high}. In this paper, we used empirical action value due to its simplicity and effectiveness. We represent actor and state value(critic) as the deep neural network. The hyper-parameter information is listed in section 5.
\subsection{Types of Actor-Critic Algorithms}
There are numerous types of AC algorithms. In this paper, we focus on classical AC from  \cite{sutton2018reinforcement}, \cite{mnih2016asynchronous}(A2C) and Proximal Policy Optimization(PPO) \cite{schulman2017proximal} which is a special version of Trust Region Policy Optimization(TRPO)\cite{schulman2015trust} that can be easily implemented. The difference between the classical policy gradient and PPO is that PPO uses a proxy objective function as the replacement of $\log\pi_\theta(a_t | s_t)$ in \ref{eq:1}. 

\begin{equation}\label{eq:2}
    maximize \quad{\mathbb{E}[\frac{\pi_\theta(a_t | s_t)}{\pi_{\theta_{old}}(a_t|s_t)}A_t}]
\end{equation}
\begin{equation}
    subject\quad to \quad {\mathbb{E}[KL[\pi_{old}(.|s_t), \pi_\theta(.|s_t)]] \leq \delta}
\end{equation}

The classical policy gradient method improves the policy directly within the parameter space, whereas PPO refines the policy within the policy space itself. This distinction leads to PPO offering greater stability and a more consistent, monotonic improvement during training \cite{schulman2015trust}. Value-based algorithms are highly developed and have demonstrated strong performance in various applications \cite{mnih2015human}. However, the challenge of decomposing the multi-agent utility function in mixed-agent environments remains unresolved due to the complexity of agent interactions. On the other hand, Actor-Critic methods, which are capable of handling both continuous and discrete action spaces, are particularly effective in environments with continuous action spaces where value-based approaches often encounter difficulties.

Both A2C and PPO are on-policy algorithms, meaning they cannot leverage past experiences for learning, which limits their sample efficiency. However, on-policy algorithms offer the advantage of mitigating non-stationarity issues in multi-agent settings, as all agents act based on their most recent policies. Despite this benefit, training Actor-Critic methods can be challenging due to their sensitivity to hyperparameters, making the tuning process critical and potentially difficult.

\subsection{Multi-Agent Actor Critics}
Multi-Agent Actor Critics is a direct extension of the Actor-Critic to multi-agent settings. One naive solution is to train each agent independently without considering the interaction nature among the agents namely independent learning (IL) \cite{claus1998dynamics}, \cite{tan1993multi}, \cite{de2020independent}. One major drawback of IL is that it suffers non-stationarity due to each agent treats other agents as part of the training environment while ignoring the evolving of the other agent's policy. As a result, independent learning approaches may produce unstable learning and may not converge to any solution.
\subsection{Centralized Training and Decentralized Execution}
An enhancement to the approach is to train a centralized critic in conjunction with a decentralized policy. In this paper, we model this framework as a decentralized partially observable Markov decision process (DEC-POMDP) with shared rewards. This structure allows for decentralized decision-making by the agents while benefiting from centralized value assessment through the critic \cite{yu2022surprising}. A DEC-POMDP is defined by $\langle S,A,O,R,P,n,\gamma \rangle.$ \textit{S} is the state space. \textit{A} is the shared action space for each agent \textit{i}. \textit{$o_i$} is the local observation for agent \textit{i}. $P(s^{'}|s,\textbf{a})$ denotes the transition probability from s to $s^{'}$ given the joint action \textbf{a} = ($a_1$,...$a_n$) for all n agents. R(s,\textbf{a}) denotes the shared reward function. \textit{$\gamma$} is the discount factor. Agents use a local policy $\pi_{\theta}((a_i|o_i)$ parameterized by $\theta_{i}$ to emit an action $a_i$ from the local observation $o_i$, and jointly optimize the discounted accumulated reward.
\[
\mathbb{E}[\sum_{t=0}^{T}\gamma^tR(s^{t}, \textbf{a}^{t})]
\]
where $\textbf{a}^{t} = (a^{t}_1,...,a^{t}_N)$ is the joint action at time step t.

This paradigm tackles the non-stationarity during training still maintaining the decentralised fashion namely centralized training and decentralized execution(CTED). \cite{lowe2017multi} provides a general Actor-Aritic framework which is suited to both cooperative and mixed environments. It works with both homogeneous and heterogeneous agents. We can extend policy gradient in equation \ref{eq:1} to centralized learning with 
\begin{equation}\label{eq:4}
    g^{t}_{i}:=\mathbb{E}[\triangledown_{\theta_i}\log\pi_{i}(a^{t}_{i}|o^{t}_{i})Q^{\pi}_{i}(x^{t},a^{t}_{1},..., a^{t}_{N})]
\end{equation}
where $o_{i}$ is the local observation of agent \textit{i}. $Q^{\pi}_{i}(x^{t},a^{t}_{1},..., a^{t}_{N})$ is the centralized version of the action value function. It takes input $\textit{x}^t$ which can be the concatenation of all the agent's local observations and $\textbf{x}=(x^{t}_{1},...,x^{t}_{N})$ as the inputs. Equation \ref{eq:4} usually can be replaced by the advantage version:

\begin{equation}\label{eq:5}
    g^{t}_{i}:=\mathbb{E}[\triangledown_{\theta_i}\log\pi_{i}(a^{t}_{i}|o^{t}_{i})A^{t}] 
\end{equation}
\[
where \quad A^{t} = Q^{\pi}_{i}(x^{t},a^{t}_{1},..., a^{t}_{N})-V^{\pi}_{i}(x^{t})
\]

$A_t{}$ here becomes centralized advantage function as well as for MAPPO. In this paper, we explore both A2C and PPO in the CTED paradigm assuming partial observability for each agent in a cooperative environment in which all the agents share a common reward. 
\subsection{Practical Algorithm and Implementation Details}
Algorithm \ref{alg_ac} provides a pseudocode for multi-agent Actor-Critic implementation. It first initializes the actor and critic for each agent. The actor is only conditional on agent's local observation. The critic for each agent takes the state which could be the concatenation of each agent's local observation as input. Given the initial state, each agent acts according to its own policy conditional on its local observation. After this joined action is applied to the environment, new state is returned from environment. Each agent receives a common reward. When the buffer is full, actor is updated with policy gradient while critic is updated by minimizing the loss between value function and target value function. For PPO, algorithm follows the suit. The difference is that PPO does not use policy gradient, rather it utilizes proxy function \ref{eq:2} to update the policy in the policy space.

\begin{algorithm}[H]
\caption{On Policy Multi-Agent Actor-Critic Algorithm}\label{alg_ac}
\begin{algorithmic}
\State Initialize $\pi_{i}(\theta_{i})$ and $\textit{V}_{i}(\phi_{i})$  for each agent \textit{i}
\State Initialize buffer \textit{D}
    \For {episode = 1, M}
    \\\:\:\quad  Receive initial state \textit{s}, observe each $\textit{o}_{i}$
	\For {t = 1, T}
		\State Select $\textit{a}_{i}$ for each agent \textit{i} according to $\pi_{i}(\textit{a}|\textit{o};\theta_{i})$
		\State Execute actions a =($\textit{a}_1,...,\textit{a}_N$), observe reward r and new state $\textit{s}^{\prime}$
            \State Store (\textit{s}, \textit{a}, \textit{r}, $\textit{s}^{\prime}$) in buffer \textit{D}
            \State \textit{s} $\gets \textit{s}^{\prime}$
            \If{ \textit{D} is full}
                \For {agent $\textit{i}=1$ to N}
                \State set $\textit{y}_{i} = \textit{r}_{i}+\gamma\textit{V}_{i}
                (\textit{s}^{\prime};\phi_{i})$
                \State set $\delta{{i}} = \textit{y}_{i} - \textit{V}_{i}
                (\textit{s}; \phi_{i})$
                \State Update critic by minimizing the loss:
                \[
                \zeta(\phi_{i})=\frac{1}{|\textit{D}|}\sum_{j=1}^{|\textit{D}|}(y_{i}^j - \textit{V}_{i}^{j}(\textit{s};\phi_{i}))^2
                \]
                
                \State Update actor using the policy gradient:
                \[
                \triangledown_{\theta_i}\xi= \frac{1}{|\textit{D}|}\sum_{j=1}^{|\textit{D}|}\triangledown_{\theta_i}\log\pi_{\theta_i}(a_{i}^j | o_{i}^{j})\delta_{i}^{j}
                \]
                \EndFor
                \State Clear buffer \textit{D}
            \EndIf
	\EndFor
    \EndFor
\end{algorithmic} 
\end{algorithm}

\section{Related Work}

\cite{wiebe2023learning} explored both independent Deep Q-Learning(IDQ) and QMIX \cite{sunehag2017value} in a cooperative setting. QMIX is a enhanced version in the sense that it decomposes overall central action value Q into multiple non-negative Local Q values respective to each agent. It compared IDQ and QMIX across three scenarios: Confidentiality, Integrity, Availability. The results showed QMIX outperformed IDQ due to its ability to consider more information when computing a central Q value while IDQ only consider each agent's local observations.

\cite{contractor2024learning} utilized Differentiable Inter-Agent Learning(DIAL) \cite{foerster2016learning} to explore the effectiveness of the communication in the multi-agent CybORG backed environment. \cite{sukhbaatar2016learning} went a step further to model the communication among the agents. In this study, the author compared DIAL with and without communication and QMIX in several phases. It showed the DIAL with communication outperformed its counter-partner DIAL without communication. It also performs comparable to QMIX due to the utilization of the fully observed environment in the critic.
Although given the successful usage of the above algorithms, they are lack of scalability due to the nature of the value-based algorithms. These value-based algorithms need to pre-define a fixed set of agents. Action selection is based on the individual agent's value. This prevent large scale deployment. To close this gap, we explore using AC base algorithms since AC provides a more general RL learning framework. It fits the generalized policy iteration paradigm which is fundamental to nearly all RL learning algorithms. Secondly it can be naturally applied to both continuous and discrete action spaces by only adjusting the actor outputs. With stochastic discrete policy, the actor needs to output a categorical distribution and takes a sample from it. With continuous actions space, the actor outputs a Gaussian distribution and takes a sample of it or through reparameteration trick \cite{kingma2013auto}. Actor also can outputs a deterministic vector of real values with continues action space \cite{silver2014deterministic}. Lastly Actor-critic is more scalable during training by applying parameter sharing when facing the homologous agent environments. 
\section{Autonomous Cyber Defense Environment}
\begin{figure}
\centering 
\includegraphics[width=0.5\textwidth]{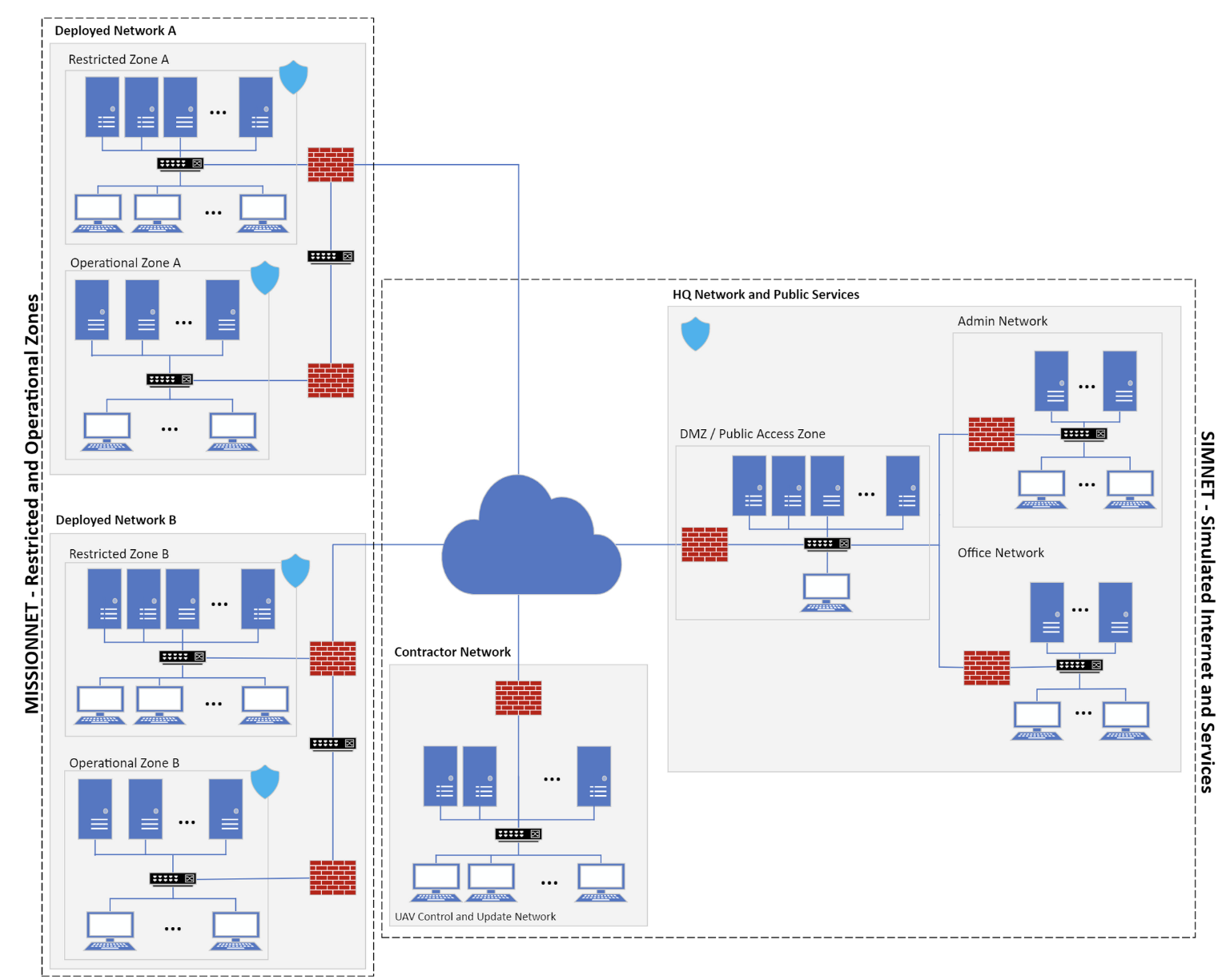}
    \caption[The system.]{Network Structure. }
\label{CAGE4}
\end{figure}

We leverage CybORG environment \cite{standen2021cyborg} known through a series of CAGE Challenges. Challenge4\cite{cage_challenge_4_announcement} is the most recent one. It provides a multi-agent partially observable cooperative environment. The network structure is seen in Fig \ref{CAGE4}. This network provides a realistic environment that can be either emulated using Amazon Web Services (AWS) or simulated with Python. The network is divided into four segments: two deployed networks, a Headquarters (HQ) network, and a Contractor network. Each deployed network is structured into two security zones: a restricted zone and an operational zone. The Headquarters network is segmented into three security zones: a Public Access Zone, an Admin Zone, and an Office Network. These networks are interconnected via the internet.

Three types of agents operate within the networks: red agents, green agents, and blue agents. Green agents function as regular users, while red agents act as attackers. Our primary focus is on developing blue agents, whose role is to defend the network against penetration by red agents. There are five blue agents in total. Each deployed network contains two blue agents, one for each security zone, and the Headquarters network is protected by a single blue agent responsible for all zones.

The red team initiates operations with access to a random machine within the Contractor Network and attempts to move laterally throughout the network. With each turn, there is a small chance that a red agent will spawn if a green agent opens a phishing email or accesses a compromised service. Each zone can host a maximum of one red agent, though these agents can maintain a presence across multiple hosts. While the blue team may successfully eliminate all traces of red agents from a network, red agents will always retain a foothold within the Contractor Network.

The primary objective of the agents is to ensure the continuity of routine operations while preventing malicious activities across the network. CAGE4 categorizes the operations into three phases, each with distinct priorities. Our focus will be on the first phase: General Operations and Maintenance.

\subsection{Observations}
CybORG returns raw observations in the form of a python dictionary which provides information related to the previous actions undertaken by both the red and blue agents. The dictionary include the details pertaining to each host such as network subnets, processes and system details. Each agent only be able to observe their local subnet information and have no knowledge of other subnets'. CAGE4 challenge provides wrappers that support MARL algorithms by conforming the raw observation to the PettingZoo Environment. PettingZoo is one popular multi-agent gym environments that can be utilized by Deep Reinforcement Learning algorithms. The observation space is MultiDiscrete. During training, they are converted into real value vectors by one-hot encoding which makes them suitable input to neural network. 

\subsection{Actions}
Blue agents can perform a series discrete actions in order to eliminate the threats from the red agents. The actions are listed in table \ref{actions}. 
\begin{table}[h]
\begin{minipage}{.5\textwidth} 
\begin{center}\small
    \begin{tabular}{ | p{1.5cm} | p{6cm} | }
    \hline
    \textbf{Actions} & \textbf{Description of action} \\ \hline
    Monitor & Collection of information about flagged malicious activity on the network. Corresponds to action ID 1: Scan in the OpenC2 specification. This action occurs automatically as a default action. \\ \hline

    Analyze & Collection of further information relating to malware, on a specific host to enable blue to better identify if red is present on the system. Corresponds to action ID 30: Investigate in the OpenC2 specification. \\ \hline
 
    DeployDecoy & Setup of a decoy service (type randomly assigned) on a specified host. Green agents do not access these services, so any access is an unambiguous indicator of red activity. Decoy services cannot replace legitimate services. \\\hline    

    Remove & Attempting to remove red from a host by destroying malicious processes, files and services. \\ \hline
    
    Restore & Restoring a system to a known good state. This has significant consequences for system availability. \\ \hline

    BlockTraffic & Allow traffic to and from the specified zone. \\ \hline
    
    AllowTraffic & Block traffic to and from the specified zone. If green agents are attempting to communicate to that zone this will result in penalties. \\ \hline
 
    \end{tabular}
    \caption {Available Actions for Blue Agents.}
    \label{actions}
\end{center}
\end{minipage}%
\end{table}
\subsection{Rewards}
Blue agents start with zero points. Negative rewards/penalties are received when green agents are not able to perform their work, when they access a compromised service, and when red agents chooses the Impact action.  All rewards for Phase 1 are shown in Table \ref{rewards}

\begin{table}[h]
\begin{center}
\begin{tabular}{ c c c c}

\begin{minipage}{.5\textwidth}
\begin{center}\small
    \begin{tabular}{ | p{1.5cm} | p{1.5cm} | p{1.5cm} |  p{1.5cm} | }
    \hline
    \textbf{Zone} & \textbf{Local Work Fails} & \textbf{Access Service Fails} & \textbf{Red impact/access} \\ \hline

    HQ Network & -1	& -1 & -3 \\ \hline
    Contractor Network	& 0	    & -5	& -5 \\ \hline
    Restricted Zone A	& -1	& -3	& -1 \\ \hline
    Operational Zone A	& -1	& -1	& -1 \\ \hline
    Restricted Zone B	& -1	& -3	& -1 \\ \hline
    Operational Zone B	&-1	& -1  & -1 \\ \hline
    Internet	& 0	& 0	& 0 \\ \hline
    \end{tabular}
    \caption {Rewards for Green Action Failures and Compromise }
    \label{rewards}
\end{center}
\end{minipage}%
\end{tabular}
\end{center}
\end{table}

\section{Result and Discussion}
We conducted two multi-agent Actor-Critic in discrete action spaces. One is classical AC implementation A2C. Another one is PPO. We explored both independent and centralized experiments for both algorithms namely independent Actor-Critic(IAC), multi-agent Actor-Critic(MAAC), Independent PPO(IPPO) and  multi-agent Actor-Critic(MAPPO). For both MAAC and MAPPO. We assume each agent only has its local observation. During training, we train centralized critics that take as input the concatenation of all agent's local observation. Each agent takes action only conditional on its local observation. To make it more a general case, each agent has it own critic. Also, we did not perform parameter sharing among the agents. This allows the algorithms take heterogeneous agents where each agent may have different observation spaces and action spaces. We also perform action masking since at each interaction, some of the actions are not valid. We hope action masking can help the actor perform more effectively. More information for action masking can be found in \cite{huang2020closer}.

The hyperparameter setup is as in table \ref{params}. To be fair for comparison, all the models have two hidden layers with size of 64. When testing different learning rate, we found Actor-Critic performed better with larger learning rate. This may indicate the model need to be updated more frequently. So we decided the model was updated every 50 steps while in both IPPO and MAPPO the model was updated every 100 steps. We used clip of 0.2 suggested from the PPO original paper \cite{schulman2017proximal}. In the experiment, each algorithm has 25 runs. Each run has 1000 episodes. Each episode has the maxim length of 50 steps. Then the average total return for each episode over 25 runs was calculated. 

\begin{figure}
\centering 
\includegraphics[width=0.4\textwidth]{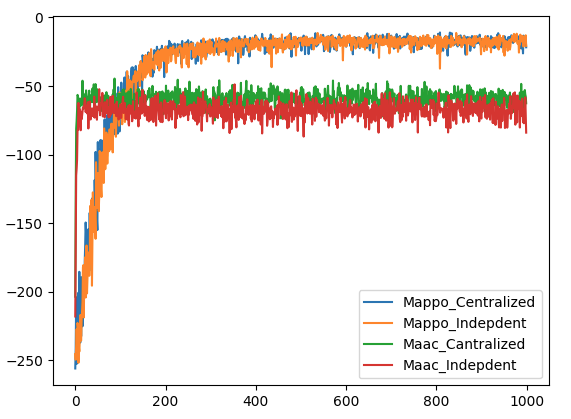}
    \caption[The system.]{Model Performance. }
\label{traning1}
\end{figure}

The model training results are shown in Fig \ref{traning1}. We see IPPO and MAPPO perform similarly. This reflects the fact IPPO usually provides a strong baseline in multi-agent environment due to its stability and efficiency. Its on policy nature can naturally mitigate the non-stationarity challenge in the multi-agent environments. Also, the agents in CAGE4 environment did not conduct explicit coordination among them to be successful. The segregation of agents into subnets limits their ability to influence the state of other agents. Those agent did not need frequent and intense physical interactions as those in MPEs from PettingZoo. This may further reduce the effectiveness of the centralized critics. Meanwhile, we see centralized MACC outperforms IAC. This shows centralized training indeed improves the effectiveness in multi-agent settings. Also, we see the MACC converges early that MAPPO. This is due to both two IAC and MAAC algorithms used larger learning rate. We test multiple learning rates in IAC and MAAC training. Smaller learning rate did not converge while large learning rate did. The reason may be that the agents did not learn quick enough to adapt the ever evolving non-stationary environment with the smaller learning rate. Larger learning rate makes the agent quickly learn the knowledge from the interaction with environment and act accordingly. But IPPO and MAPP was trained with much smaller learning rate, both of them still converge at the reasonable time. It may be due to that PPO learns in the policy space rather than parameter space. This makes PPO agents quickly adjust their policy up to the new environment after each updating. Lastly, we see both MAPP implementations outperform MAAC's implementation in both centralized and independent training. This reflects the fact that PPO as trust region base Actor-Critic has superior performance comparing to classic policy gradient based implementation of Actor-Critic algorithms due to its clipped surrogate objective function.

\begin{table}[h]
\begin{minipage}{.5\textwidth}
\begin{center}\small
    \begin{tabular}{ | p{2cm} | p{2cm} | p{2cm} |}
    \hline
    \textbf{Parameters} & \textbf{Actor-Critic} & \textbf{PPO} \\ \hline

    Network & [64. 64]	 & [64, 64] \\ \hline
    Leaning Rate	& 0.05	& 1e-4 \\ \hline
    $\gamma$ 	& 0.99	& 0.95	\\ \hline
    Clip    & 	& 0.2	\\ \hline
    Steps to Update & 50	& 100  \\ \hline
    Activation	& ReLU	& ReLU	\\ \hline
    Optimizer	& Adam	& Adam	\\ \hline  
    
    \end{tabular}
    \\
    \caption {Hyperparameters }
    \label{params}
\end{center}
\end{minipage}%

\end{table}

\section{Conclusion}
We explored both A2C and PPO in partially observable multi-agent cooperative setting in cyber defense systems. Previous studies are more focused on value-based which is lack of scalability, Our work explored policy-based algorithms. Policy-based algorithms provide a solution which is more scalable. We provided detailed implementation of both MAAC, MAPPO, IAC and IPPO. We demonstrated both A2C and PPO can effectively learn from the interactions with the simulated environment through both IL and CTDE paradigms. MAPPO performs similarly in both training paradigms. AC performs better in CTDE than in IL. This is due to the less frequently interactions among the agents. In summary, we conclude both algorithms can learn efficiently in the multi-agent settings.


\ifCLASSOPTIONcompsoc
  \section*{Acknowledgments}
\else
  \section*{Acknowledgment}
\fi
The authors would like to thank AiDN(AI-Defined Networking) team provide invaluable support on this project.



%




\bibliographystyle{ieeetr}
\bibliography{references}

\end{document}